\documentclass[
 amsmath,amssymb,
 aps,twocolumn,
]{revtex4-2}

\usepackage{graphicx}
\usepackage{dcolumn}
\usepackage{bm}
\usepackage{hyperref}
\usepackage{cleveref}
\usepackage{mathtools}

\begin{document}

\title{A fresh look at boundary terms in Einstein-Hilbert gravity via an initial value variational principle}

\author{Songmin Ha}
 \email{hyh8679@korea.ac.kr}

\author{Alexander Rothkopf}%
 \email{akrothkopf@korea.ac.kr}
 
\affiliation{Department of Physics, Korea University, Seoul 02841, Republic of Korea}

\date{\today}

\begin{abstract}
A key tenet of general relativity is the dynamical nature of space-time, ideally represented as an initial value problem. Here we explore the variational formulation of classical Einstein-Hilbert gravity as initial value problem by constructing its Schwinger-Keldysh-Galley (SKG) action, including a careful treatment of boundary terms. The construction is based on a doubling of degrees of freedom and independent of a foliation. The action naturally decomposes into a bulk term furnishing Einstein's equations and a boundary term, which is related to conserved quantities, such as the Komar mass. We find that since only trivial connecting conditions must be specified on boundaries, the variational action principle for gravity as an initial value problem is rendered well-posed without the need to add additional boundary terms. The SKG approach to gravity offers a novel and complementary avenue to solve for the metric of spacetime directly from the action, bypassing the governing equations.
\end{abstract}

\maketitle


\section{Introduction}

The general theory of relativity is a key ingredient in the understanding of dense celestial bodies, galaxies and the structure of the universe as a whole (for textbooks see e.g.\ \cite{zeldovich1971relativistic, weinberg2008cosmology, straumann2012general, choquet2015introduction}). It has proven to accurately capture phenomena on vastly different energy scales, from the minute ripples of space-time in the form of gravitational waves to the inescapable gravitational wells produced by black holes (for a recent review see e.g.\ \cite{gupta2025yearsextremegravitytests}). The modern study of these phenomena relies heavily on computational methods (see e.g.\ \cite{sperhake2015numerical}), which are based on the 3+1 decomposition of spacetime and the ADM Hamiltonian approach (reviewed in  \cite{Gourgoulhon:2007ue}). One sets out to solve the resulting constrained Cauchy problem on the level of governing equations. General covariance, partially hidden due to the choice of a foliation is recovered through the inclusion of a set of primary and secondary constraints, the latter being the so-called Hamiltonian and momentum constraints.

At the same time that Einstein constructed the governing equations of gravity, Hilbert developed an action formulation of general relativity \cite{hilbert1915grundlagen}. Hilbert's approach follows the conventional boundary value problem setting of Hamilton's variational principle. It proposes an action (or score) functional $S_{\rm EH}$ which depends on the metric, its first and second derivatives. If the metric is known and fixed on an initial spacelike hypersurface and crucially is also known and fixed at a final spacelike hypersurface, then the extremum of $S_{\rm EH}$ produces the classical metric of the system realized in nature, equivalent to the solution of Einstein's equations (For a discussion of the Weiss boundary value variational principle see \cite{feng_weiss_2018}).  

In the 1970's York \cite{york_role_1972} as well as Gibbons and Hawking \cite{gibbons_action_1977} carefully studied the boundary terms associated with the conventional action formulation of gravity. They realized that since the Einstein-Hilbert action contains second derivatives of the metric, its variation leads to boundary terms, which only vanish if both the metric, as well as its covariant derivative are set to zero on the boundary. At first sight this may not sound problematic, as in fact, a scalar theory action may be written either with $\frac{1}{2}(\partial_\mu\phi)^2$ or $-\frac{1}{2}\phi\Box \phi$. Boundary terms that make reference to $\phi$ and $\partial \phi$ can be dealt with by imposing Dirichlet or von Neumann conditions. However in gravity the metric and its covariant derivative are not independent, but related via intrinsic constraints. In electromagnetism there is only a single such constraint, i.e. Gauss' law, but for gravity there are four constraints that lead to a nontrivial relation between the metric and its derivatives. The accepted solution to this conundrum is to add by hand a boundary term to the Einstein-Hilbert action, the GHY term, which cancels with the offending boundary term during variation.

Here we take a different perspective, related to the modern variational treatment of initial value problems, known as the Schwinger-Keldysh-Galley (SKG) approach (see \cite{Galley:2012hx} for first work in point mechanics, see \cite{rothkopf2024unifying} for an application to constrained systems and see \cite{Rothkopf:2024hxi} for an application to field theory). Starting point is the realization that the need to impose acausal boundary data (on the final spacelike hypersurface), a key drawback of Hamilton's principle, prevents the use of that action principle to actually predict the evolution of a dynamical system. The occurrence of acausal boundary terms can be avoided however by going over to a formulation of the action in terms of a set of doubled degrees of freedom, distributed on a forward and backward branch. Assigning the Lagrangian of the original theory to the forward branch d.o.f. and minus the original Lagrangian to the backward branch d.o.f. one constructs an action whose boundary terms under variation exhibit advantageous properties. Instead of having to specify the value and derivatives of the d.o.f. on the future boundary---knowledge not available to us---one only has to ensure that the values and derivatives on the forward and backward branch agree. These so-called connecting conditions do not require the imposition of external information and thus causality is not affected.

The quantum doubled degrees-of-freedom formulation is well established. Its classical limit has e.g.\ been deployed in Ref. \cite{salcedo_open_20252} to develop an open-systems formulation of gravity, providing novel insights into the effects of dark matter and dark energy on the propagation of gravitational waves. The role of boundary terms however was not explored in that work.

In this study we construct the SKG action for Einstein-Hilbert gravity, paying particular attention to boundary terms. We find that only trivial connecting conditions need to be imposed to make boundary terms vanish and in turn the variational principle for gravity as an initial value problem emerges as well defined. As we will show in detail, the SKG action decomposes into a bulk term furnishing Einstein's equations and a boundary term, which is related to conserved quantities.

We proceed as follows: We briefly review Hamilton's action principle for gravity in \cref{sec:HamiltonEH} including the origin of the GHY term. In \cref{sec:SKG} we introduce the Schwinger-Keldysh-Galley action principle to initial value problems before constructing the SKG action for Einstein-Hilbert gravity in \cref{sec:SKGEH}. We study the Noether current of SKG gravity in \cref{sec:Noether} before concluding in \cref{sec:Concl}.

\section{Hamilton's principle for Einstein-Hilbert gravity}
\label{sec:HamiltonEH}

The dynamical degrees of freedom of general relativity are represented by the metric tensor $g_{\mu\nu}$ of a spacetime $\cal{M}$. Hilbert's action deploys the Ricci scalar, the only invariant measure of space-time curvature linear in the second derivatives of the metric (there are none that only contain first order derivatives) \cite{vermeil1917notiz}. This scalar is formed ultimately from the affine connection of spacetime, the Christoffel symbols
\begin{align}
    \Gamma^{\lambda}{}_{\mu\nu} = \frac{1}{2} g^{\lambda\rho}
\left( \partial_\mu g_{\rho\nu} + \partial_\nu g_{\rho\mu} - \partial_\rho g_{\mu\nu} \right),
\end{align}
combined into the Riemann curvature tensor
\begin{align}
    R^{\rho}{}_{\sigma\mu\nu}= \partial_\mu \Gamma^{\rho}{}_{\nu\sigma}- \partial_\nu \Gamma^{\rho}{}_{\mu\sigma}+ \Gamma^{\rho}{}_{\mu\lambda}\Gamma^{\lambda}{}_{\nu\sigma}- \Gamma^{\rho}{}_{\nu\lambda}\Gamma^{\lambda}{}_{\mu\sigma}.
\end{align}
The Ricci scalar $R$ is but an appropriate contraction of the Riemann tensor $R^{\rho}{}_{\sigma\mu\nu}$ into the Ricci tensor $R_{\mu\nu}$ as follows
\begin{align}
    R = g^{\mu\nu} R_{\mu\nu} = g^{\mu\nu} R^{\lambda}{}_{\mu\lambda\nu}.
\end{align}
The Einstein-Hilbert action of gravity thus reads
\begin{align}
    S_{EH} = \frac{1}{16\pi G}\int d^4 x \sqrt{-g}~R,
\end{align}
where $G$ is the gravitational constant and $\sqrt{-g}= \sqrt{-\rm{det}[g]}$ renders the integral measure invariant under diffeomorphisms. 

In order to obtain the classical equations of motion for gravity (in the absence of matter) Hamilton's principle suggests to vary the action w.r.t. the components of the metric tensor. Following the rules of variational calculus one obtains \cite{lim}:
\begin{align}
    \nonumber &\delta S_{EH} = \\
    &\frac{1}{16\pi G}\int_{\cal{M}} d^4x\sqrt{-g}~G_{\mu\nu} \delta g^{\mu\nu} + \frac{1}{16\pi G}\int_{\partial \cal{M}} d^3x \sqrt{\gamma}n_\rho V^\rho\label{eq:varEH},
\end{align}
where the first integral is carried out over the bulk of spacetime $\cal M$. The second integral denotes a boundary term carried out over hypersurfaces $\partial \cal{M}$, which are characterized by their normal four-vector $n_\rho$. $\gamma$ denotes the induced metric on the boundary and the integrand reads 
\begin{align}
    V^\rho = \nabla^\rho \delta g^\lambda{}_\lambda - \nabla_\lambda \delta g^{\rho \lambda} \label{eq:bndint}.
\end{align}
The covariant derivative applied to a two-tensor yields $\nabla_\mu \delta g^{\alpha\beta}
= \partial_\mu \delta g^{\alpha\beta}
+ \Gamma^{\alpha}{}_{\mu\lambda} \delta g^{\lambda\beta}
+ \Gamma^{\beta}{}_{\mu\lambda} \delta g^{\alpha\lambda}$, while it reduces to a simple partial derivative when applied to a scalar $\nabla_\mu \delta g^\lambda{}_\lambda = \partial_\mu \delta g^\lambda{}_\lambda$.

All the ingredients for a successful application of Hamilton's principle appear to be in place. Locating the metric, whose variations leave the action unchanged $ \delta S_{\rm EH}=0$ will yield the vacuum Einstein governing equations
\begin{align}
    G_{\mu\nu} = R_{\mu\nu} - \frac{1}{2} g_{\mu\nu} R=0,
\end{align}
if the boundary terms can be made to vanish. But it is here that Hamilton's principle asks us to specify acausal data. To make \cref{eq:bndint} vanish we must set the variations of $g$ and its covariant derivative to zero on a future spacelike hypersurface. However we do not know apriori what the solution of the metric is, we are trying to solve for it in the first place. On top of that the variation of the metric and its covariant derivative are not independent but are related via the intrinsic Hamiltonian and momentum constraints. Thus choosing appropriate boundary conditions even on timelike hypersurfaces constitutes a highly non-trivial problem.

To avoid having to deal with these complications Gibbons, Hawking and York suggested to add to the Einstein-Hilbert action a boundary term by hand, which under variation produces the negative of the boundary term in \cref{eq:varEH}, leading to the standard expression (used e.g. in \cite{Brown_1993,Brown_19932})
\begin{align}
    S_{\rm EH+ GHY} = \frac{1}{16\pi G}\int d^4 x \sqrt{-g}~R +\frac{1}{8}\int_{\partial M}d^3x \sqrt{\gamma} n^\mu{}_{;\mu}.
\end{align}

The goal of this paper is to explore an alternative to this treatment by considering an action approach specifically geared towards initial value problems, the Schwinger-Keldysh-Galley variational principle. It avoids the need to provide acausal boundary data altogether by introducing a doubling of degrees of freedom.

\section{Schwinger--Keldysh--Galley action principle}
\label{sec:SKG}

In preparation for its application to gravity, we briefly review here the continuum Schwinger-Keldysh-Galley (SKG) formalism for the simplest case of point particle mechanics and its extensions to scalar field theory.

The SKG formalism amounts to a variational action principle that is causal and hence only makes reference to initial data.
The method originates from the Schwinger--Keldysh closed-time-path
formalism developed for non-equilibrium quantum systems
\cite{Schwinger1961,Keldysh1964}, and was independently rediscovered in the context of classical point mechanics
by Galley in an attempt to formulate a consistent variational principle for
non-conservative systems \cite{Galley:2012hx}.
The key idea is to double the degrees of freedom by introducing two
independent histories of the system and to construct an action that
compares the difference in their dynamical evolution.

\subsection{Point particle mechanics}

For a point particle whose trajectory we wish to describe by the function $x(t)$, we may construct the boundary value action of Hamilton's principle
\begin{align}
S_{\mathrm{BVP}}
&=
\int_{t_i}^{t_f} dt\,
L\bigl[x(t), \dot{x}(t)\bigr],
\end{align}
where $L[x,\dot x]$ denotes the classical Lagrangian of the system. Following \cite{Galley:2012hx}, let us instead consider two copies of the trajectory functions $x_1(t)$ and $x_2(t)$ and make the following assignment of Lagrangians with opposite relative sign
\begin{align}
S_{\mathrm{IVP}}
&=
\int_{t_i}^{t_f} dt\,
\Bigl(
L\bigl[x_1(t), \dot{x}_1(t)\bigr]
-
L\bigl[x_2(t), \dot{x}_2(t)\bigr]
\Bigr)\\
&=
\int_{t_i}^{t_f} dt\,
L\bigl[
x_1,\dot x_1,
x_2,\dot x_2
\bigr].
\end{align}
Both trajectories here are integrated forward in time and the labels $x_{1/2}$ do not imply time reversal. 

It turns out that the physics of the initial value formulation is most lucidly displayed when considering the following linear combination of trajectories
\begin{align}
x_+ = \frac{1}{2}(x_1 + x_2),
\qquad
x_- = x_1 - x_2 ,
\end{align}
where $x_+$ is referred to as the retarded and $x_-$ as the advanced component.

Varying the action leads to bulk and boundary terms which explicitly read
\begin{align}
\nonumber&\delta S_{\mathrm{IVP}}[x_+,\dot x_+,x_-,\dot x_-]
=\\
&\int_{t_i}^{t_f} dt\,
\Bigg(
\Bigl[
\frac{\partial L}{\partial x_+}
-
\frac{d}{dt}
\frac{\partial L}{\partial \dot{x}_+}
\Bigr]\delta x_+
+
\Bigl[
\frac{\partial L}{\partial x_-}
-
\frac{d}{dt}
\frac{\partial L}{\partial \dot{x}_-}
\Bigr]\delta x_-
\Bigg)
\nonumber\\
&\quad
+
\left[
\frac{\partial L}{\partial \dot{x}_+}\,\delta x_+
+
\frac{\partial L}{\partial \dot{x}_-}\,\delta x_-
\right]_{t_i}^{t_f}.\label{eq:SKGppvar}
\end{align}
Similar to the situation in Hamilton's principle we have to deal with boundary terms at the initial and final time slice. However unlike Hamilton's principle we now have additional freedom to make them vanish. 

Imposing initial conditions on $x_+$ and $\dot x_+$ takes care of the boundaries at initial time. While it may appear that one must impose a condition on $x_-$ at initial time, a careful derivation of the SKG approach from the discretized quantum mechanical path integral reveals that imposition of $\dot x_+$ is intimately related to correctly treating $x_-$ at initial time (for details see \cite{horowitz2026}). 

At final time, we find that setting $x_-$ to zero removes one of the boundary terms. Note that this choice did not fix the value of $x_1$ or $x_2$ but merely states that their values have to agree. In a causal initial value problem, we cannot fix $x_+$ at final time. However the boundary term making reference to $\delta x_+$ includes $\partial L/\partial \dot x_+=\pi_-$, which for standard kinetic terms quadratic in the coordinates evaluates to $\pi_- \propto \dot x_-$. Thus by requiring $\dot x_-=0$ we can remove also the second boundary term at final time without the need to specify acausal data. The fixing of $x_-$ and $\dot x_-$ at final time is referred to as \textit{connecting conditions}.

With the variation of the retarded and advanced components being independent, the terms multiplying $\delta x_+$ and $\delta x_-$ in \cref{eq:SKGppvar} encode the equations of motion for the advanced and retarded components respectively. When deriving the SKG formalism from the path integral it becomes clear that only terms linear in the advanced components contribute as the imposition of the connecting conditions renders $x_-(t)=0$ everywhere when the variation of the action vanishes. Intuitively it follows from the fact that taking a derivative w.r.t. $x_+$ produces equations of motion for $x_-$ which are at least linear in $x_-$ and thus allow for the $x_-=0$ solution which emerges once its value and derivative have been fixed at final time. The emergence of $x_-=0$ is intimately connected to the classical limit of the SKG approach from the quantum path integral and is called the physical limit. If ultimately $x_-(t)=0$ it is clear that only terms that are linear in the advanced components contribute to the e.o.m. for the retarded components generated by variation w.r.t. $x_-$.

\begin{figure*}
\includegraphics[scale=0.5]{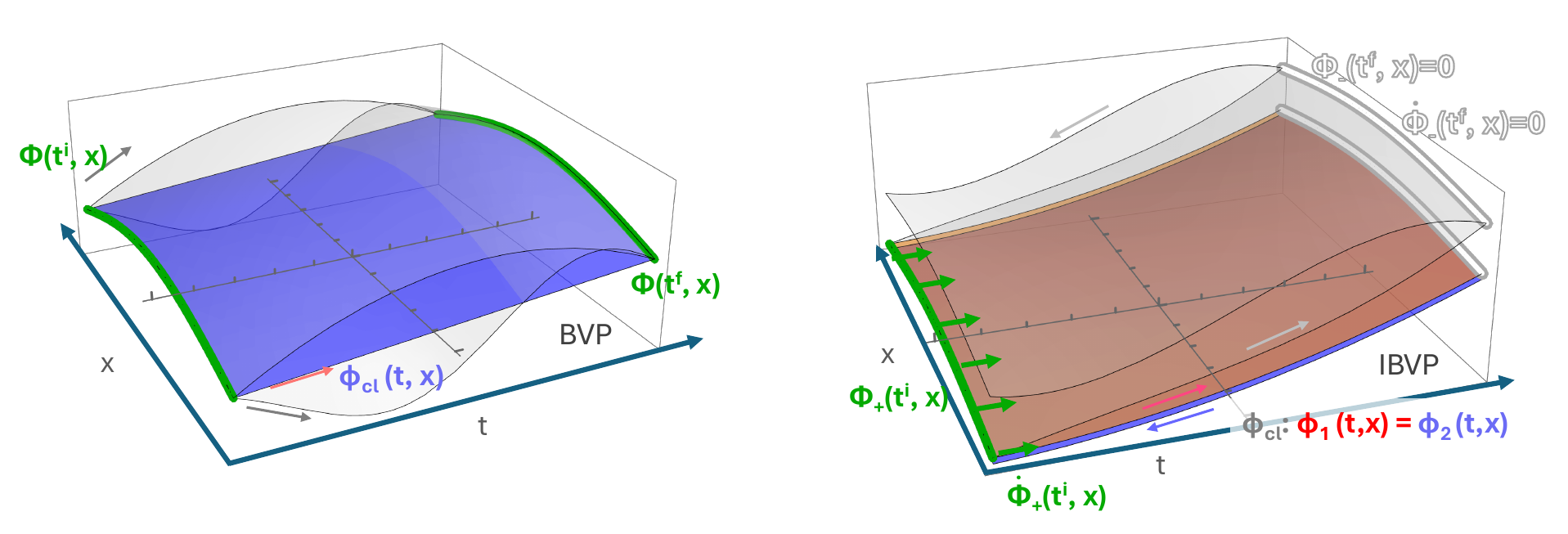}
\caption{(left) Hamilton's boundary value action principle for field theory requires specification of field data at initial and final time (green). After imposition of these Dirichlet temporal boundary values the classical field configuration (blue) emerges from the extremum of the action $S_{\rm BVP}$ under arbitrary variations (gray). (right) The SKG action principle asks us to specify only initial data for the physical degrees of freedom $\phi_+(t_i,x)$ and $\dot \phi_+(t_i,x)$ (green). The advanced components $\phi_-(x)$ and their temporal derivatives are set to zero on the final time slice (white). The extremum of the action under imposition of initial and connecting conditions produces the classical field configuration for the retarded components $\phi_{\rm cl}(x)=\phi_+(x)=\big(\phi_1(x)+\phi_2(x)\big)/2$ for which the solution on the forward and backward branch agrees such that $\phi_-(x)=\phi_1(x)-\phi_2(x)=0$.}
\label{fig:sketch}
\end{figure*}

We may thus succinctly state the SKG action principle to obtain the classical trajectory $x_{\rm cl}$ via
\begin{align}
\left. \frac{\delta S_{\rm IVP}}{\delta x_-}\right|_{x_-=0,x_+=x_{\rm cl}}=0.
\end{align}

As described in detail in \cite{Rothkopf:2022zfb} the formulation of an initial value action opens novel avenues to obtain the classical trajectory of the point particle. After discretizing the first derivatives in the action by e.g.\ a finite difference scheme, such as summation-by-parts, one carries out a numerical optimization in the doubled dynamical degrees of freedom $x_+$ and $x_-$ locating the extremum of the discretized action. This procedure reveals the classical trajectory without the need to derive the equations of motion of the system of interest. 

Note that the numerical implementation on the action level forces us to make explicit the external data supplied. Initial and connecting conditions are conveniently included in the action via additional Lagrange multiplier terms. In contrast to Hamilton's action principle, Lagrange multipliers affecting the retarded $x_+$ degrees of freedom are only needed at initial time. At final time Lagrange multipliers enforce the connecting conditions setting $x_-$ and its time derivative to zero. Being formulated directly in terms of an integral, solving for the trajectory on the action level offers access to weak solutions inaccessible directly from ordinary differential equations, such as the conventional governing equations (see e.g. \cite{rothkopf2024unifying}).

Let us close this subsection with a brief discussion of symmetry structure. The doubling of the degrees of freedom has apparently enlarged the symmetry space of the theory. We may consider transformations which act in the same way on the forward and backward branch or, which act in opposite direction on the branches. If acting on the system leaves the action intact and amounts to a global continuous transformation then it may appear that the number of associated Noether charges too doubles. As discussed in e.g.\ \cite{sieberer_keldysh_2016,rothkopf2024unifying} this is not the case as in the physical limit only a single Noether charge, that associated with opposite direction transformations, survives. The other apparent Noether charge vanishes identically as the advanced components are set to zero. I.e. the physical symmetry space of the original theory is maintained in the retarded sector irrespective of the intermediate doubling of the degrees of freedom.

\subsection{Scalar field theory}

The structure of the SKG formalism generalizes straightforwardly
from point particle mechanics to classical scalar field theory (for a sketch of the setup see \cref{fig:sketch}). Instead of a single trajectory $x(t)$, the dynamical variable is now a scalar field $\phi(x)$ defined over flat spacetime with coordinates $x^\mu= (t,x,y,z)^\mu$ with $\mu\in\{0,1,2,3\}$.

For a real scalar field with Lagrangian density
$\mathcal{L}[\phi,\partial_\mu \phi]$, the standard boundary value
action reads
\begin{align}
S_{\mathrm{BVP}}
=
\int d^4x \,
\mathcal{L}\bigl[\phi(x),\partial_\mu\phi(x)\bigr].
\end{align}
As before, Hamilton's principle requires fixing the variations at the
initial and final times, as well as on the spatial boundaries.

In the SKG formulation we instead introduce two independent field
configurations $\phi_1(x)$ and $\phi_2(x)$ and define the action involving the doubled degrees of freedom
\begin{align}
S_{\mathrm{IBVP}}
=
\int d^4x \,
\Big(
\mathcal{L}[\phi_1,\partial_\mu \phi_1]
-
\mathcal{L}[\phi_2,\partial_\mu \phi_2]
\Big).
\end{align}
Both fields are defined on the same spacetime manifold and are treated
as independent dynamical variables (for construction of the approach to scalar wave propagation see \cite{Rothkopf:2024hxi}).

As in the mechanical case, it is convenient to introduce
average and difference fields,
\begin{align}
\phi_+ = \frac{1}{2}(\phi_1+\phi_2),
\qquad
\phi_- = \phi_1-\phi_2,
\end{align}
that we will refer to as the retarded and the advanced fields respectively.

Variation of the SKG field theory action yields
\begin{align}
&\delta S_{\mathrm{IBVP}}
=
\int d^4x \,
\Big(
E_+ \,\delta \phi_+
+
E_- \,\delta \phi_-
\Big)
+\\
&\sum_{a=1}^3\left[\int dt d^2x \big[ \ldots \big]\right]_{x^a_i}^{x^a_f} +
\left[\int d^3x \big[ \ldots \big]\right]_{t_i}^{t_f},
\end{align}
where
\begin{align}
E_\pm
=
\frac{\partial \mathcal{L}}{\partial \phi_\pm}
-
\partial_\mu
\frac{\partial \mathcal{L}}{\partial (\partial_\mu \phi_\pm)}.
\end{align}

In contrast to the mechanical case we now have to deal with boundaries in both temporal and spatial directions arising from derivatives in the different spacetime directions $\partial_\mu\phi_{\pm}$ inside the SKG action. 

The treatment of the temporal boundaries proceeds entirely analogous to the mechanical case. At the initial spacelike hypersurface defined by $t=t_i$ one fixes the physical initial data,
\begin{align}
    \phi_+(t_i,x^a)=\phi_0(x^a)\quad \partial_t \phi_+(t_i,x^a)=\dot \phi_0(x^a).
\end{align}
Note that by prescribing $\phi_+(t_i,x^a)$ at initial time one has already provided all initial spatial derivatives too. The boundary terms at the final spacelike hypersurface, i.e. at $t=t_f$, are taken care of by \textit{connecting conditions}, setting the advanced components to zero
\begin{align}
        \phi_-(t_f,x^a)=0\quad \partial_t \phi_-(t_f,x^a)=0.
\end{align}

The spatial boundary terms are either avoided all together by prescribing periodic boundary conditions or can be set to zero by appropriately assigning Dirichlet or von Neumann boundary conditions to the retarded components, while setting to zero the advanced ones.  

Inspecting the terms $E_\pm$ arising from the variation of the retarded and advanced components one again finds that the variation w.r.t. to the retarded components $\delta\phi_+$ yields the equation of motion for $\phi_-$. It admits the trivial vanishing solution, which emerges by imposing the connecting conditions at final time. The variation of the advanced components  $\delta\phi_-$ on the other hand leads to the equations of motion for the retarded components, which are nothing but the Euler Lagrange equations for the classical field $E_+$. We may thus formulate a similarly compact stationarity condition for the scalar field theory SKG action
\begin{align}
\left. \frac{\delta S_{\rm IBVP}}{\delta \phi_-}\right|_{\phi_-=0,\phi_+=\phi_{\rm cl}}=0.
\end{align}

Similar to the point particle case, the field theory  action for initial boundary value problems offers a novel route to solving for the classical field configuration. As laid out in the context of scalar wave propagation in \cite{Rothkopf:2024hxi} we can discretize the derivatives in the various spacetime directions using summation-by-parts finite difference operators. Now in addition to specifying initial and connecting conditions in temporal direction via Lagrange multipliers, one also needs to introduce Lagrange multipliers on the spatial boundaries if these are to be treated explicitly. Solving for the extremum of the discretized action then reveals the classical field configuration without the need to derive the equations of motion. 

Note that in the context of a gauge theory the SKG action allows to formulate the initial boundary value problem without the need to impose a choice of gauge. This is in contract to the Hamiltonian approach, where the equations of motion emerge after fixing of a gauge.

The question of symmetries in the SKG formalism for field theory is of particular interest in the case of gauge theories. For a detailed analysis and lucid exposition in the case of electromagnetism, we refer the reader to \cite{salcedo_open_2025}. The upshot is that the apparent doubling of the symmetry space at intermediate steps in the physical limit is reduced to the original symmetry space present in the retarded components. The imposition of connecting conditions on the advanced components is closely connected with this reduction.

\section{Initial-value action principle for gravity}
\label{sec:SKGEH}

With the SKG formalism for field theory at hand, we are ready to construct an initial value action for Einstein-Hilbert gravity and investigate the role of boundary terms.

\subsection{Constructing the SKG Einstein-Hilbert action}

Starting point is the introduction of two copies of the metric
\begin{align}
g_{(1)\mu\nu},\qquad g_{(2)\mu\nu},
\end{align}
considered here simply as the two dynamical fields of our theory. It is not necessary to invoke the concept of a forward and backward time branch at this point, the two copies live in all of spacetime ${\cal M}$. In anticipation of the fact that an SKG action is most cleanly formulated in the retarded $(+)$ and advanced $(-)$ components we define 
\begin{align}
&g_{\mu\nu} \equiv g_{(+)\mu\nu} = \frac{1}{2} \left( g_{(1)\mu\nu}+g_{(2)\mu\nu} \right)\\
&a_{\mu\nu} \equiv g_{(-)\mu\nu} = g_{(1)\mu\nu}-g_{(2)\mu\nu}.
\end{align}
Conversely we may express
\begin{align}
g_{(1)\mu\nu}=g_{\mu\nu}+\frac{1}{2}a_{\mu\nu},
\qquad
g_{(2)\mu\nu}=g_{\mu\nu}-\frac{1}{2}a_{\mu\nu}.
\end{align}

As we saw in the previous section, after variation and taking the physical limit, the retarded components of the theory represent the classical degrees of freedom. It therefore is natural to associate the covariant derivative $\nabla_\rho$ with the Levi-Civita connection compatible with the retarded metric $g_{\mu\nu}$
\begin{align}
\nabla_\rho g_{\mu\nu}=0.
\end{align}
For notational convenience let us define the quantities
\begin{align}
a \equiv g^{\mu\nu}a_{\mu\nu},
\qquad 
a^{\mu\nu} := g^{\mu\alpha}g^{\nu\beta}a_{\alpha\beta}.
\end{align}
The reader may worry that the doubling of the degrees of freedom has modified the symmetry structure of the theory. We refer to the excellent discussion of the apparent extended symmetry space and its reduction to the physical one in the Schwinger-Keldysh approach in \cite{salcedo_open_20252}.

The advanced components $a_{\mu\nu}$ vanish in the physical limit and thus need only be treated to linear order in the action. Correspondingly all functions of the doubled metric tensors can be expanded to linear order in $a_{\mu\nu}$.The first instance where such an expansion becomes relevant is the computation of the inverse metric tensors. Let us define the mixed tensor $H_\mu{}^\nu = g_{\mu\rho}a^{\rho\nu}$ and consider the following Neumann series
\begin{align}
(g\pm\tfrac{1}{2}h)^{-1}
= g^{-1}\Big(I\pm\tfrac{1}{2}H\Big)^{-1}
= g^{-1}\sum_{n=0}^{\infty}\Big(\mp \tfrac{1}{2}\Big)^n H^n,
\end{align}
to linear order in $a$ we obtain
\begin{align}
g_{(1)}^{\mu\nu}=g^{\mu\nu}-\frac{1}{2}a^{\mu\nu}+\mathcal{O}(a^2),
\;
g_{(2)}^{\mu\nu}=g^{\mu\nu}+\frac{1}{2}a^{\mu\nu}+\mathcal{O}(a^2).
\end{align}
It is important to note that no information relevant for the determination of the extremum of the SKG action has been lost by truncating the expansion at second order in $a$. Using the series expansion of the matrix logarithm, the determinant 
\begin{align}
    \sqrt{\det(A+B)}
= \sqrt{\det(A)}\,\exp\!\left[
\frac{1}{2}\sum_{n=1}^{\infty}
\frac{(-1)^{n+1}}{n}\,
\operatorname{tr}\!\bigl((A^{-1}B)^n\bigr)
\right].
\end{align}
becomes
\begin{align}
\sqrt{-g_{(1)}}
&=\sqrt{-g}\left[
1+\frac{1}{4}a
+\mathcal{O}(a^2)
\right],
\\
\sqrt{-g_{(2)}}
&=\sqrt{-g}\left[
1-\frac{1}{4}a
+\mathcal{O}(a^2)
\right].
\end{align}

What is still missing for the construction of the SKG action is the decomposition of the Ricci scalar into retarded and advanced components. Since we only need to keep track of contributions linear in $a_{\mu\nu}$, it may be treated similarly to a perturbation on top of the background field $g_{\mu\nu}$. This means we can reuse a standard result from perturbation theory, describing the variation of the Ricci tensor w.r.t. a perturbation $a$
\begin{align}
\delta R_{\mu\nu}[a]
=\frac{1}{2}\Big(
\nabla_\rho\nabla_\mu a^\rho{}_\nu
+\nabla_\rho\nabla_\nu a^\rho{}_\mu
-\Box a_{\mu\nu}
-\nabla_\mu\nabla_\nu a
\Big).
\label{eq:linricci}
\end{align}
We stress that no perturbative expansion has taken place as terms of higher order in $a$ do not contribute to the extremum of the action in the SKG approach.

Now we are ready to construct the full SKG action for Einstein-Hilbert gravity. It is obtained from assigning the conventional Lagrangian to the two copies of the metric with opposite relative sign
\begin{align}
\mathcal{L}_{\text{SKG}}
\equiv \sqrt{-g_{(1)}}\,R[g_{(1)}]-\sqrt{-g_{(2)}}\,R[g_{(2)}].
\end{align}
The decomposition of the determinant and Ricci tensor above tells us that 
\begin{align}
&R_{\mu\nu}(g_{(1)}) = R_{\mu\nu}[g]+\frac{1}{2}\delta R_{\mu\nu}[a]+\mathcal{O}(a^2),\\
&R_{\mu\nu}(g_{(2)}) = R_{\mu\nu}[g]-\frac{1}{2}\delta R_{\mu\nu}[a]+\mathcal{O}(a^2),
\end{align}
so that for the retarded and advanced components we find
\begin{align}
&R_{\mu\nu}(g_{(1)})+R_{\mu\nu}(g_{(2)})=2R_{\mu\nu}[g]+\mathcal{O}(a^2)\\
&R_{\mu\nu}(g_{(1)})-R_{\mu\nu}(g_{(2)})=\delta R_{\mu\nu}[a]+\mathcal{O}(a^2).
\end{align}

Keeping only terms up to order $\mathcal{O}(a)$, one finds
\begin{widetext}
\begin{align}
\mathcal{L}_{\text{SKG}}
&=\sqrt{-g}[g^{\mu\nu}(R_{\mu\nu(1)}-R_{\mu\nu(2)})-\frac{1}{2}a^{\mu\nu}(R_{\mu\nu(1)}+R_{\mu\nu(2)})+\frac{1}{4}ag^{\mu\nu}(R_{\mu\nu(1)}+R_{\mu\nu(2)})]\\
&= \sqrt{-g}\Big(
g^{\mu\nu}\,\delta R_{\mu\nu}[a]
-a^{\mu\nu}R_{\mu\nu}[g]
+\frac{1}{2}a\,R[g]
\Big)
+\mathcal{O}(a^2).
\label{eq:LskO1}
\end{align}
\end{widetext}

Let us go one step further and take a closer look at the term containing $\delta R_{\mu\nu}$. It turns out that it can be rewritten as a boundary term. Starting from \eqref{eq:linricci} and using metric compatibility,
\begin{align}
&g^{\mu\nu}\,\delta R_{\mu\nu}[a]\label{eq:dividend}\\
\nonumber &=\frac{1}{2}\Big(
2\nabla_\rho\nabla_\mu a^{\rho\mu}
-\Box a
-\Box a
\Big)
=\nabla_\rho\Big(\nabla_\mu a^{\rho\mu}-\nabla^\rho a\Big).
\end{align}
Since we end up with a covariant derivative acting on a vector field we may use the covariant Stokes theorem (see appendix E of \cite{carroll2019spacetime}) to write 
\begin{align}
\int_{\mathcal{M}} d^4x\,\sqrt{-g_r}\,\nabla_\rho V^\rho
=\int_{\partial\mathcal{M}} d^3x\,\sqrt{|\gamma|}\,n_\rho V^\rho,
\end{align}
which produces the boundary contribution
\begin{align}
S_{\text{bdy}}
=\int_{\partial\mathcal{M}} d^3x\,\sqrt{|\gamma|}\,
n_\rho\Big(\nabla_\mu a^{\rho\mu}-\nabla^\rho a\Big).
\label{eq:SKGBoundary}
\end{align}
In full we find the following SKG action for Einstein-Hilbert gravity
\begin{widetext}
\begin{align}
\mathcal{S}_{\text{SKG}}
&= \int_{\cal M} d^4x \sqrt{-g}\Big(
-a^{\mu\nu}R_{\mu\nu}[g]
+\frac{1}{2}a\,R[g]
\Big) + \int_{\partial \cal M} d^3x ,\sqrt{|\gamma|}\,
n_\rho\Big(\nabla_\mu a^{\rho\mu}-\nabla^\rho a\Big)
\label{eq:SKGEinsteinHilbert}.
\end{align}
\end{widetext}
This action naturally decomposes into a bulk term over ${\cal M}$ and a boundary term over $\partial \cal M$, each of which makes reference to the retarded $g_{\mu\nu}$ and advanced $a_{\mu\nu}$ components of the metric. 

\subsection{Variation and boundary term analysis}

Let us derive the equations of motion of the dynamical degrees of freedom of the theory $g_{\mu\nu}$ and $a_{\mu\nu}$ by variation of the action we just derived. In the bulk, variations of $a$ do not generate any boundary terms, as no derivatives act on the advanced metric. The variation w.r.t. $g$ produces bulk terms with a complicated dependence on $g$ but all of them depend linearly on $a_{\mu\nu}$. Since the Ricci scalar $R[g]$ contains derivatives of the retarded metric, their variation also produce boundary terms, which are linear in the advanced metric too.

Noting that the variation of the trace of the advanced metric is $\delta a = a^{\mu\nu}\,\delta g_{\mu\nu} + g_{\mu\nu}\,\delta a^{\mu\nu}$ we find
\begin{widetext}
\begin{align}
\nonumber \delta S_{\mathrm{SKG}}
=
-\int_{M} d^4x \,\sqrt{-g}\,
\left(R_{\mu\nu} - \tfrac{1}{2} g_{\mu\nu} R\right)
\delta a^{\mu\nu}
\;+\;
\int_{\partial M} d^3x \,\sqrt{|\gamma|}\, n_\rho
\left(\nabla_\mu \delta a^{\rho\mu} - \nabla^\rho (g_{\alpha\beta}\,\delta a^{\alpha\beta})\right)\\
+ \int_{\cal M} d^4x \sqrt{-g}\Big( a_{\mu\nu} f[g] + a \tilde f_{\mu\nu}[g] \Big)\delta g_{\mu\nu} \;+\;
\int_{\partial M} d^3x \,\sqrt{|\gamma|}\, n_\rho
\left( a \bar f[\delta g^{\mu\nu}] + a_{\mu\nu}\bar{\bar f}^{\mu\nu}[\delta g^{\mu\nu}] \right)
.\label{eq:varSKG}
\end{align}
\end{widetext}
Let us inspect first the bulk terms of \cref{eq:varSKG}, which will furnish the relevant equations of motion. Since $g_{(1)}$ and $g_{(2)}$ and thus ultimately $a$ and $g$ are independent of each other they also may be varied independently. Thus there are no restrictions on neither the choice of $\delta a^{\mu\nu}$ nor $\delta g^{\mu\nu}$. In the absence of boundary contribution the first bulk term thus establishes Einstein's vacuum equations in terms of the retarded metric $g_{\mu\nu}$. This is consistent with the understanding that in the SKG action principle the classical equations of motion arise in the retarded sector from variation with the advanced components.

The second bulk term thus must establish the equations of motion for the advanced metric $a_{\mu\nu}$. In the physical limit the advanced components must be taken to zero, i.e. the equations of motion must admit the vanishing solution $a_{\mu\nu}=0$. Irrespective of the complicated dependence of the second bulk term on the retarded metric $g$ (abbreviated in the functions $f$ and $\tilde f_{\mu\nu}$) we find that the equation of motion is linear in $a$, which always admits the necessary $a_{\mu\nu}=0$ solution.

Now let us turn to the two boundary terms. The one that originated in the naturally occurring boundary term of the SKG action contains reference to the variations of the advanced metric. The boundary term arising from derivatives in the Ricci tensor on the other hand makes reference to the variation of the retarded metric but remains linear in $a_{\mu\nu}$. It is here that the SKG formalism can play out its strength. We understand that causality prevents us from knowing the solution of the retarded metric at the final spacelike hypersurface. I.e. we cannot set $\delta g_{\mu\nu}$ to zero by Dirichlet boundary conditions in the future. In addition the boundary contributions on timelike hypersurfaces require an intricate tuning of $\delta g_{\mu\nu}$ and its covariant derivatives due to the presence of intrinsic constraints. The SKG formalism allows us to avoid these difficulties, as we only need to make statements about the advanced metric $a_{\mu\nu}$, which in the physical limit vanishes. I.e. we know the appropriate boundary conditions everywhere, which shall enforce $a_{\mu\nu}=0$ and the vanishing of all derivatives. This is achieved by imposing the so-called connecting conditions of the SKG formalism
\begin{align}
a_{\mu\nu}\big|_{\partial M} = 0,
\qquad
\nabla_\rho a_{\mu\nu}\big|_{\partial M} = 0\label{eq:conconEH}.
\end{align}
Since $a_{\mu\nu}$ must only fulfill the trivial vanishing solution we are free to choose the boundary conditions (similarly to us choosing connecting conditions for the scalar field in \cref{sec:SKG}). 

Imposing boundary conditions on gauge fields directly, immediately invites the question of gauge dependence. Since our connecting conditions are formulated purely in terms of the advanced components they do not affect the retarded sector of the theory, i.e. diffeomorphism invariance of $g_{\mu\nu}$ remains intact. \Cref{eq:conconEH} explicitly breaks general diffeomorphism invariance of the advanced sector. This is however benign as it is understood (see the discussion in \cite{salcedo_open_20252}) that in the Schwinger-Keldysh formalism it is 4d retarded diffeomorphism invariance, which remains as the physical symmetry space in the physical limit.

With the connecting conditions identified, we have completed the construction of the Schwinger-Keldysh-Galley action principle for Einstein-Hilbert gravity. It provides a consistent formulation of gravity as initial value problem on the action level. Variation of the SKG action and taking the physical limit produces the correct classical equations of motion for the retarded components.

We would like to stress that the SKG approach opens up a novel avenue for obtaining the metric of spacetime, via direct optimization of the SKG action functional, without the need to derive and solve the governing equations.

As it has been understood that boundary terms in general relativity are intimately related to conserved quantities, let us explore how this connection arises in the SKG formalism in the next section.

\section{SKG action and conserved quantities}
\label{sec:Noether}

General relativity is unique, since one of the physical quantity that produces curvature, mass, is equivalent to energy. In the field theory of matter and interactions in the standard model the charge of the theory is distinct from energy.

In electromagnetism the total charge contained in a spatial volume is identified through Gauss' law. In its integral form, it asks us to inspect the electric field on the boundary in relation to the surface that encloses the spatial volume. I.e. inspecting the behavior of the system on the boundary reveals the charge content in the interior. Interestingly in electromagnetism it turns out that the quantity we are asked to inspect on the boundary, the electric field, when squared represents the local energy density inside the volume. I.e. information about charge and mass are naturally contained within the electric field. It is not clear whether a similar quantity exists for gravity.

In general relativity mass can be intuitively understood as a charge, generating curvature of spacetime. In analogy with electromagnetism it is therefore not surprising that there exists a quantity, which, when inspected on a boundary hypersurface will tell us about the mass content of the bulk. It is with this intuition that we embark on inspection of the boundary term in the SKG action.

\subsection{Noether currents}

At its core, Noether's theorem in field theory tells us to inspect the boundary terms generated by infinitesimal transformations in our system. If these transformations represent a symmetry of the action, we are led to quantities, which (if appropriate spatial boundary conditions are fulfilled) exhibit the same value on the initial and final time boundary. For an insightful discussion of Noether's theorem in general relativity in the context of Hamilton's principle see \cite{haro_noethers_2021} and the classic Ref.~\cite{iyer1994some}. 

In the SKG formalism the doubling of the degrees of freedom implies an enlarged symmetry space (for a detailed discussion see e.g. \cite{salcedo_open_20252}). In particular it is possible to act with  diffeomorphisms independently on each of the copies $g_{(1)}$ and $g_{(2)}$. Still, our task is to inspect the boundary terms generated by these infinitesimal transformations in order to identify possible conserved quantities. Note that the doubling of the degrees of freedom does not lead to a doubling of the conserved charges. This results from the fact that only those conserved quantities that are solely formulated in the retarded contributions can take on non-vanishing values in the physical limit (see \cite{salcedo_open_20252} and \cite{rothkopf2024unifying}).

Let us consider translations in field theory in flat spacetime first. Introducing a constant shift $x^\mu \to x^\mu \pm \xi^\mu$ can be represented as Taylor expansion of the fields $\phi(x)\to \phi(x) \mp \xi^\mu \partial_\mu \phi(x)$. I.e.\ the infinitesimal transformation amounts in the action to a replacement of field degrees of freedom. In the SKG formalism we may consider applying the transformation in opposite directions on $\phi_{(1)}$ and $\phi_{(2)}$ so that $\delta \phi_{(+)} = \frac{1}{2} \xi^\mu\partial\phi_{(-)}$ and $\delta \phi_{(-)} = 2 \xi^\mu\partial\phi_{(+)}$.

In general relativity we must consider not constant translations but local ones generated by a vector field $\xi^\mu(x)$ as $x^\mu \to x^\mu \pm \xi^\mu(x)$. We may think of applying $\xi^\mu_1(x)$ to $g_{(1)}$ and $\xi^\mu_2(x)$ to $g_{(2)}$. For notational convenience let us define the quantities 
\begin{align}
\xi_r^\mu:=\frac12(\xi_1^\mu+\xi_2^\mu),
\qquad
\xi_a^\mu:=\xi_1^\mu-\xi_2^\mu.
\end{align}

Since we are interested in boundary terms that survive the physical limit we here focus on transformations that act in opposite direction on the two copies of the metric. I.e. we consider a single vector field $\xi^\mu(x)$ and apply shifts such that $\xi_r^\mu=0$ and $\xi_a^\mu=2 \xi^\mu(x)$. Since local translations in general relativity are simply another set of diffeomorphisms and tensors change under diffeomorphism according to the Lie derivative ${\cal L}$ we obtain
\begin{align}
\delta_{\xi_a} g_{\mu\nu} &= \frac12\,\mathcal L_{\xi_a}a_{\mu\nu},
&
\delta_{\xi_a} a_{\mu\nu} &= 2\,\mathcal L_{\xi_a}g_{\mu\nu}.
\label{eq:offdiag_diffeo}
\end{align}
Note the peculiar structure of these transformations, where the variation of $g_{\mu\nu}$ produces a term involving $a_{\mu\nu}$ and vice versa. In particular we obtain for the variation of the advanced metric
\begin{align}
\delta_{\xi_a}a_{\mu\nu}=2\,\mathcal L_{\xi_a}g_{\mu\nu}
=2(\nabla_\mu\xi_{a\nu}+\nabla_\nu\xi_{a\mu}).
\end{align}

From our previous result \cref{eq:varSKG} we see that variation of the bulk term of the SKG action produces the equations of motions which we assume to be fulfilled. The boundary term involving $\delta g_{\mu\nu}$ vanishes in the physical limit. I.e. the only surviving contribution arises from the original boundary term of the SKG action
\begin{align}
\nonumber &\delta_{\xi}S_{\rm SK}
=\\
&\int_{\partial\mathcal M} d^3x\,\sqrt{|\gamma|}\,
n_\rho\,
\underbracket{\Big[
2\Box\xi_a^\rho
-2\nabla^\rho(\nabla_\mu\xi_a^\mu)
+2R^\rho{}_{\sigma}\xi_a^\sigma
\Big]}_{J^\rho}.
\label{eq:offdiag_boundary_current}
\end{align}
We have now identified the relevant boundary terms, as intended by Noether's theorem. The integrand evaluated on the different hypersurfaces of the boundary represents the different components of the associated Noether current $J^\mu$. In order to identify a conserved quantity we must evaluate this boundary term for a true symmetry of the action. This is the case if the vector field $\xi^\mu(x)$ is represented by a Killing vector of the theory. In that case no matter whether we translate forward or backward $\delta S=0$. In the simple case of an absence of spatial boundaries, \cref{eq:offdiag_boundary_current} tells us that the quantity in square brackets when integrated over the initial spacelike hypersurface takes on the same value as when integrated over the final spacelike hypersurface. In this specific sense it constitutes a conserved quantity.

The fact that the conserved current arises from the naturally occurring boundary term in the SKG approach mirrors closely the picture of Brown and York \cite{Brown_1993}. There conserved quantities are understood as surface integrals of geometric data on the boundary hypersurface. The key difference is that in a genuine initial value problem formulation no data on the final spacelike hypersurface needs to be supplied apriori.

\subsection{Relation to the Komar mass}

It is known that spacetimes with a timelike Killing vector admit the definition of the Komar mass \cite{komar1959covariant} as conserved quantity (see chapter 6 of \cite{carroll2019spacetime}). To shed light on the relation of our boundary term, i.e. the Noether current of \cref{eq:offdiag_boundary_current} and the concept of Komar current let us inspect the terms in the square brackets in \cref{eq:offdiag_boundary_current}.

From the relation between covariant derivatives and the curvature tensor acting on vector fields
\begin{align}
[\nabla_\mu,\nabla_\nu] \xi^\rho
=
R^\rho{}_{\sigma\mu\nu} \xi^\sigma 
\end{align}
and contraction of indices we obtain the general identity
\begin{align}
\nabla_\nu\nabla^\rho \xi^\nu
=
\nabla^\rho(\nabla_\nu\xi^\nu)+R^\rho{}_{\sigma}\xi^\sigma .
\end{align}
The definition of the Komar two-form reads
\begin{align}
K_a^{\rho\nu}:=\nabla^\rho\xi_a^\nu-\nabla^\nu\xi_a^\rho.
\end{align}
Together it allows us to write the boundary term as
\begin{align}
\nonumber &\delta_{\xi_a}S_{\rm SK}
=\\
&\int_{\partial\mathcal M} d^3x\,\sqrt{|\gamma|}\,
n_\rho\,
\Big[\underbracket{\nabla_\nu K_a^{\rho\nu}}_{J_K^\rho}+4R^\rho{}_{\sigma}\xi_a^\sigma
\Big].
\label{eq:offdiag_boundary_current}
\end{align}
which reveals that the Noether current identified in the SKG formalism is related to the Komar current $J_K^\rho$ amended by a curvature dependent term $4R^\rho{}_{\sigma}\xi_a^\sigma$. 

For the explicit example of the Schwarzschild metric the Ricci tensor vanishes identically so that the above Noether current agrees with the Komar current, which in turn produces the Komar mass as conserved quantity.

Let us point out that the SKG Noether current containing the Komar two-form amended by a curvature contribution closely mirrors the structure found in the conventional study of Noether currents in the context of gravity as boundary value problem \cite{iyer1994some}.  This similarity implies that the SKG construction equally reflects its geometric nature akin to the Iyer–Wald charge, while implementing a genuine initial-value formulation of EH gravity. 

\section{Conclusion}
\label{sec:Concl}

We have presented the construction of the Schwinger-Keldysh-Galley action for Einstein-Hilbert gravity. Particular care was taken to keep track of boundary contributions. We find that the SKG action, expressed in terms of the retarded $g_{\mu\nu}$ and advanced $a_{\mu\nu}$ components of the metric naturally decomposes into a bulk term and a boundary term. 

Variation of the SKG action reveals the classical equations of motion for the retarded components and equations of motion for the advanced components that admit the trivial $a_{\mu\nu}=0$ solution realized in the physical limit. In contrast to Hamilton's variational principle we show that no acausal boundary data needs to be supplied to make the boundary terms vanish. Instead we must enforce connecting conditions which set $a_{\mu\nu}$ and its derivatives to zero on the boundary. Since the advanced components ultimately have to vanish the imposition of these connecting conditions is not impeded by nontrivial constraints.

The SKG action principle thus allows to formulate Einstein-Hilbert gravity as a well-posed initial value problem on the level of the system action. Similar to its application in point-particle mechanics and field theory, the SKG formulation of gravity opens up a novel avenue to obtaining the classical metric by direct optimization of the action functions without the need to formulate or solve the equations of motion.

The naturally arising boundary term in the SKG action of gravity is shown to be intimately related to conserved quantities. Its variation under opposite direction translation diffeomorphisms furnishes a Noether current, which we identify with the Komar current amended by a curvature dependent term. In simple spacetimes, such as Schwarzschild, which admit a timelike Killing vector for which the variation of the action vanishes, our Noether current thus defines a conserved quantity that agrees with the Komar mass.

We believe that the SKG perspective of Einstein-Hilbert gravity not only offers valuable conceptual clarity w.r.t. the treatment of boundary terms but also constitutes the first step towards developing novel reliable numerical solvers based on the direct optimization of the SKG action.

\section*{Acknowledgements}
A.\ R.\ thanks W.\ A.\ Horowitz for frutiful collaboration and illuminating insight on the boundary structure of the SKG approach. S.\ H.\ and A.\ R.\ thank Alex Bentley Nielsen for his support. A.\ R.\ gladly acknowledges support by Korea University through project K2510461 {\it Fully Dynamical Coordinate Maps for Space-Time Symmetry Preserving Lattice Field Theory} as well as project K2511131 and K2503291.

\section*{Author contributions}
A.R.\ proposed the application of the SKG formulation to EH gravity to avoid acausal boundary terms. S.H.\ carried out the construction of the SKG action including boundary terms with guidance by A.R. . S.H.\ worked out the relation of the SKG Noether current to the Komar current. Editing of the manuscript by A.R.\ based on a first draft by S.H..

\bibliography{BndGravitySK}

@PREAMBLE{
 "\providecommand{\noopsort}[1]{}" 
 # "\providecommand{\singleletter}[1]{#1}%" 
}

@book{weinberg2008cosmology,
  title={Cosmology},
  author={Weinberg, S.},
  isbn={9780191523601},
  lccn={2007049371},
  url={https://books.google.co.kr/books?id=2wlREAAAQBAJ},
  year={2008},
  publisher={OUP Oxford}
}

@book{choquet2015introduction,
  title={Introduction to General Relativity, Black Holes, and Cosmology},
  author={Choquet-Bruhat, Y.},
  isbn={9780199666454},
  lccn={2014934913},
  url={https://books.google.co.kr/books?id=1C8UDAAAQBAJ},
  year={2015},
  publisher={Oxford University Press}
}

@book{zeldovich1971relativistic,
  title={Relativistic Astrophysics, 2: The Structure and Evolution of the Universe},
  author={Zel'dovich, I.A.B. and Novikov, I.D.},
  isbn={9780226979571},
  lccn={77128549},
  series={Relativistic Astrophysics},
  url={https://books.google.co.kr/books?id=taQcCrPtg40C},
  year={1971},
  publisher={University of Chicago Press}
}

@book{straumann2012general,
  title={General Relativity},
  author={Straumann, N.},
  isbn={9789400754096},
  lccn={2012950312},
  series={Graduate Texts in Physics},
  url={https://books.google.co.kr/books?id=jjBMw0KFtZgC},
  year={2012},
  publisher={Springer Netherlands}
}

@misc{gupta2025yearsextremegravitytests,
      title={Ten years of extreme gravity tests of general theory of relativity with gravitational-wave observations}, 
      author={Anuradha Gupta},
      year={2025},
      eprint={2511.15890},
      archivePrefix={arXiv},
      primaryClass={gr-qc},
      url={https://arxiv.org/abs/2511.15890}, 
}

@preprint{Gourgoulhon:2007ue,
    author = "Gourgoulhon, Eric",
    title = "{3+1 formalism and bases of numerical relativity}",
    eprint = "gr-qc/0703035",
    archivePrefix = "arXiv",
    month = "3",
    year = "2007"
}

@article{sperhake2015numerical,
  title={The numerical relativity breakthrough for binary black holes},
  author={Sperhake, Ulrich},
  journal={Classical and Quantum Gravity},
  volume={32},
  number={12},
  pages={124011},
  year={2015},
  publisher={IOP Publishing}
}

@article{hilbert1915grundlagen,
  title={Die grundlagen der physik.(erste mitteilung.)},
  author={Hilbert, David},
  journal={Nachrichten von der Gesellschaft der Wissenschaften zu G{\"o}ttingen, Mathematisch-Physikalische Klasse},
  volume={1915},
  pages={395--408},
  year={1915}
}

@article{iyer1994some,
  title={Some properties of the Noether charge and a proposal for dynamical black hole entropy},
  author={Iyer, Vivek and Wald, Robert M},
  journal={Physical review D},
  volume={50},
  number={2},
  pages={846},
  year={1994},
  publisher={APS}
}

@article{sieberer_keldysh_2016,
	title = {Keldysh field theory for driven open quantum systems},
	volume = {79},
	issn = {0034-4885},
	url = {https://dx.doi.org/10.1088/0034-4885/79/9/096001},
	doi = {10.1088/0034-4885/79/9/096001},
	number = {9},
	urldate = {2023-01-18},
	journal = {Reports on Progress in Physics},
	publisher = {IOP Publishing},
	author = {Sieberer, L. M. and Buchhold, M. and Diehl, S.},
	month = aug,
	year = {2016},
	note = {Number: 9},
	pages = {096001},
}

@article{feng_weiss_2018,
	title = {The {Weiss} variation of the gravitational action},
	volume = {50},
	issn = {1572-9532},
	url = {https://doi.org/10.1007/s10714-018-2420-2},
	doi = {10.1007/s10714-018-2420-2},
	number = {8},
	urldate = {2026-01-26},
	journal = {General Relativity and Gravitation},
	author = {Feng, Justin C. and Matzner, Richard A.},
	month = jul,
	year = {2018},
	pages = {99},
}

@article{york_role_1972,
	title = {Role of {Conformal} {Three}-{Geometry} in the {Dynamics} of {Gravitation}},
	volume = {28},
	copyright = {http://link.aps.org/licenses/aps-default-license},
	issn = {0031-9007},
	url = {https://link.aps.org/doi/10.1103/PhysRevLett.28.1082},
	doi = {10.1103/PhysRevLett.28.1082},
	number = {16},
	urldate = {2026-02-01},
	journal = {Physical Review Letters},
	author = {York, James W.},
	month = apr,
	year = {1972},
	pages = {1082--1085},
}

@article{gibbons_action_1977,
	title = {Action integrals and partition functions in quantum gravity},
	volume = {15},
	copyright = {http://link.aps.org/licenses/aps-default-license},
	issn = {0556-2821},
	url = {https://link.aps.org/doi/10.1103/PhysRevD.15.2752},
	doi = {10.1103/PhysRevD.15.2752},
	number = {10},
	urldate = {2025-08-18},
	journal = {Physical Review D},
	author = {Gibbons, G. W. and Hawking, S. W.},
	month = may,
	year = {1977},
	pages = {2752--2756},
}

@article{Galley:2012hx,
    author = "Galley, Chad R.",
    title = "{Classical Mechanics of Nonconservative Systems}",
    eprint = "1210.2745",
    archivePrefix = "arXiv",
    primaryClass = "gr-qc",
    doi = "10.1103/PhysRevLett.110.174301",
    journal = "Phys. Rev. Lett.",
    volume = "110",
    number = "17",
    pages = "174301",
    year = "2013"
}

@article{rothkopf2024unifying,
  title = {Variational approach to nonholonomic and inequality-constrained mechanics},
  author = {Rothkopf, A. and Horowitz, W. A.},
  journal = {Phys. Rev. E},
  pages = {--},
  year = {2026},
  month = {Jan},
  eprint = {arXiv:2409.11063},
  publisher = {American Physical Society},
  doi = {10.1103/rch9-wsbw},
  url = {https://link.aps.org/doi/10.1103/rch9-wsbw}
}

@article{Rothkopf:2024hxi,
    author = {Rothkopf, Alexander and Horowitz, W. A. and Nordstr{\"o}m, Jan},
    title = "{Exact symmetry conservation and automatic mesh refinement in discrete initial boundary value problems}",
    eprint = "2404.18676",
    archivePrefix = "arXiv",
    primaryClass = "math.NA",
    doi = "10.1016/j.jcp.2024.113686",
    journal = "J. Comput. Phys.",
    volume = "524",
    pages = "113686",
    year = "2025"
}

@article{Brown_19932,
   title={Microcanonical functional integral for the gravitational field},
   volume={47},
   ISSN={0556-2821},
   url={http://dx.doi.org/10.1103/PhysRevD.47.1420},
   DOI={10.1103/physrevd.47.1420},
   number={4},
   journal={Physical Review D},
   publisher={American Physical Society (APS)},
   author={Brown, J. David and York, James W.},
   year={1993},
   month=feb, pages={1420–1431} }

@article{Brown_1993,
   title={Quasilocal energy and conserved charges derived from the gravitational action},
   volume={47},
   ISSN={0556-2821},
   url={http://dx.doi.org/10.1103/PhysRevD.47.1407},
   DOI={10.1103/physrevd.47.1407},
   number={4},
   journal={Physical Review D},
   publisher={American Physical Society (APS)},
   author={Brown, J. David and York, James W.},
   year={1993},
   month=feb, pages={1407–1419} }

@misc{haro_noethers_2021,
	title = {Noether's {Theorems} and {Energy} in {General} {Relativity}},
	url = {http://arxiv.org/abs/2103.17160},
	doi = {10.48550/arXiv.2103.17160},
	urldate = {2025-12-01},
	publisher = {arXiv},
	author = {Haro, Sebastian De},
	month = mar,
	year = {2021},
	note = {arXiv:2103.17160 [physics]
version: 1},
}

@preprint{salcedo_open_20252,
	title = {An {Open} {System} {Approach} to {Gravity}},
	url = {http://arxiv.org/abs/2507.03103},
	doi = {10.48550/arXiv.2507.03103},
	urldate = {2025-12-13},
	publisher = {arXiv},
	author = {Salcedo, Santiago Agüí and Colas, Thomas and Dufner, Lennard and Pajer, Enrico},
	month = jul,
	year = {2025},
	note = {arXiv:2507.03103 [hep-th]},
}

@article{vermeil1917notiz,
  title={Notiz {\"u}ber das mittlere Kr{\"u}mmungsma{\ss} einer n-fach ausgedehnten Riemann'schen Mannigfaltigkeit},
  author={Vermeil, Hermann},
  journal={Nachrichten von der Gesellschaft der Wissenschaften zu G{\"o}ttingen, Mathematisch-Physikalische Klasse},
  volume={1917},
  pages={334--344},
  year={1917}
}

@article{salcedo_open_2025,
	title = {An {Open} {Effective} {Field} {Theory} for light in a medium},
	volume = {2025},
	issn = {1029-8479},
	url = {http://arxiv.org/abs/2412.12299},
	doi = {10.1007/JHEP03(2025)138},
	number = {3},
	urldate = {2025-12-18},
	journal = {Journal of High Energy Physics},
	author = {Salcedo, Santiago Agui and Colas, Thomas and Pajer, Enrico},
	month = mar,
	year = {2025},
	note = {arXiv:2412.12299 [hep-th]},
	pages = {138},
}

@book{carroll2019spacetime,
  title={Spacetime and Geometry: An Introduction to General Relativity},
  author={Carroll, S.M.},
  isbn={9781108775557},
  url={https://books.google.co.kr/books?id=1XSmDwAAQBAJ},
  year={2019},
  publisher={Cambridge University Press}
}

@misc{lim,
  author        = {Malcolm Perry and Ian Lim},
  title         = {Lecture notes in General Relativity},
  month         = {January},
  year          = {2019},
  publisher={Department of Applied Mathematics and Theoretical Physics},
  url = {https://lim.physics.ucdavis.edu/teaching/files/gr-notes-partiii.pdf}
  
}

@article{Rothkopf:2022zfb,
    author = {Rothkopf, Alexander and Nordstr{\"o}m, Jan},
    title = "{A new variational discretization technique for initial value problems bypassing governing equations}",
    eprint = "2205.14028",
    archivePrefix = "arXiv",
    primaryClass = "math.NA",
    doi = "10.1016/j.jcp.2023.111942",
    journal = "J. Comput. Phys.",
    volume = "477",
    pages = "111942",
    year = "2023"
}

@preprint{horowitz2026,
    author = "W.\ A. Horowitz and A. Rothkopf",
    title = "(in preparation)",
    month = "3",
    year = "2026"
}

@article{komar1959covariant,
  title={Covariant conservation laws in general relativity},
  author={Komar, Arthur},
  journal={Physical Review},
  volume={113},
  number={3},
  pages={934},
  year={1959},
  publisher={APS}
}

@article{Schwinger1961,
  author = {Schwinger, J.},
  title = {Brownian Motion of a Quantum Oscillator},
  journal = {Journal of Mathematical Physics},
  volume = {2},
  pages = {407--432},
  year = {1961}
}

@article{Keldysh1964,
  author = {Keldysh, L. V.},
  title = {Diagram Technique for Nonequilibrium Processes},
  journal = {Zh. Eksp. Teor. Fiz.},
  volume = {47},
  pages = {1515--1527},
  year = {1964}
}

\end{document}